\documentclass[ %
 reprint,
superscriptaddress,
 amsmath,amssymb,
 aps,
prb,
]{revtex4-1}

\usepackage{graphicx}
\usepackage{dcolumn}
\usepackage{bm}
\usepackage{subfigure}
\usepackage{epstopdf}
\usepackage{lettrine}

\begin{document}

\preprint{APS/123-QED}

\title{Coherent Spin Amplification Using a Beam Splitter }

\author{Chengyu Yan}
\email{uceeya3@ucl.ac.uk}
\affiliation{%
	London Centre for Nanotechnology, 17-19 Gordon Street, London WC1H 0AH, United Kingdom\\
}%
\affiliation{
	Department of Electronic and Electrical Engineering, University College London, Torrington Place, London WC1E 7JE, United Kingdom
}%
\author{Sanjeev Kumar}
\affiliation{%
	London Centre for Nanotechnology, 17-19 Gordon Street, London WC1H 0AH, United Kingdom\\
}%
\affiliation{
	Department of Electronic and Electrical Engineering, University College London, Torrington Place, London WC1E 7JE, United Kingdom
}%
\author{Kalarikad Thomas}
\email{Currently at: Department of Physics, Central University of Kerala, Riverside Transit Campus, Padannakkad, Kerala 671314, India}
\affiliation{%
	London Centre for Nanotechnology, 17-19 Gordon Street, London WC1H 0AH, United Kingdom\\
}%
\affiliation{
	Department of Electronic and Electrical Engineering, University College London, Torrington Place, London WC1E 7JE, United Kingdom
}%
\author{Patrick See}
\affiliation{%
	National Physical Laboratory, Hampton Road, Teddington, Middlesex TW11 0LW, United Kingdom\\
}%
\author{Ian Farrer}
\email{Currently at: Department of Electronic and Electrical Engineering, University of Sheffield, Mappin Street, Sheffield S1 3JD, United Kingdom}
\affiliation{%
	Cavendish Laboratory, J.J. Thomson Avenue, Cambridge CB3 OHE, United Kingdom\\
}%
\author{David Ritchie}
\affiliation{%
	Cavendish Laboratory, J.J. Thomson Avenue, Cambridge CB3 OHE, United Kingdom\\
}%
\author{Jonathan Griffiths}
\affiliation{%
	Cavendish Laboratory, J.J. Thomson Avenue, Cambridge CB3 OHE, United Kingdom\\
}%
\author{Geraint Jones}
\affiliation{%
	Cavendish Laboratory, J.J. Thomson Avenue, Cambridge CB3 OHE, United Kingdom\\
}%
\author{Michael Pepper}
\affiliation{%
	London Centre for Nanotechnology, 17-19 Gordon Street, London WC1H 0AH, United Kingdom\\
}%
\affiliation{
	Department of Electronic and Electrical Engineering, University College London, Torrington Place, London WC1E 7JE, United Kingdom
}%

\date{\today}

\begin{abstract}

We report spin amplification using a capacitive beam splitter in n-type GaAs where the spin polarization is monitored via transverse electron focusing measurement. It is shown that partially spin-polarized current injected by the emitter can be precisely controlled and the spin polarization associated with it can be amplified by the beam splitter, such that a considerably high spin polarization of around 50\% can be obtained. Additionally, the spin remains coherent as shown by the observation of quantum interference. Our results illustrate that spin polarization amplification can be achieved in materials without strong spin-orbit interaction.

\end{abstract}

\maketitle

Introduction.---Controlled manipulation of electron spin has been a major area of research for developing future spin-based logic devices. The realization of such devices requires controllable spin transport to remain coherent over a long distance. The spin transistor fundamentally relies on the manipulation of electron spin rather than the current which drives the device. This can offer advantages in terms of accuracy and speed in comparison to the conventional transistor. First proposed by Datta and Das\cite{DD90}, a spin transistor utilizes the spin precession\cite{DD90,FKR99,OYB99,EBL03,BK14,CHS15,LCF17}, which can be induced by magnetic material or spin-orbit interaction (SOI) to control the transmission of a charge carrier with a particular spin orientation. However, it is difficult to extend spin transistor scheme relying on spin precession to materials  such as n-type GaAs owing to weak SOI despite advantages such as high electron mobility and long spin relaxation time. Although one can deposit magnetic material on GaAs, there is a problem with interface scattering and the magnetic materials may influence the spin distribution and could result in undesirable decoherence\cite{KSG16}. It is, thus, of broad interest to achieve a non-spin-precession approach without relying on the strong SOI.

In the present work, we demonstrate an approach to spin-polarization amplification which does not rely on the SOI. In this approach, an enhancement in controlled spin polarization in a one-dimensional (1D) channel is achieved by exploiting the transmission probability of two spin branches and quantum interference through a capacitive beam splitter erected between the emitter and collector in the transverse magnetofocusing configuration. With this approach, we can achieve a net spin polarization of $\sim$50\% (with scope of considerable improvement).

\begin{figure}
	
	\includegraphics[height=1.5in,width=3.0in]{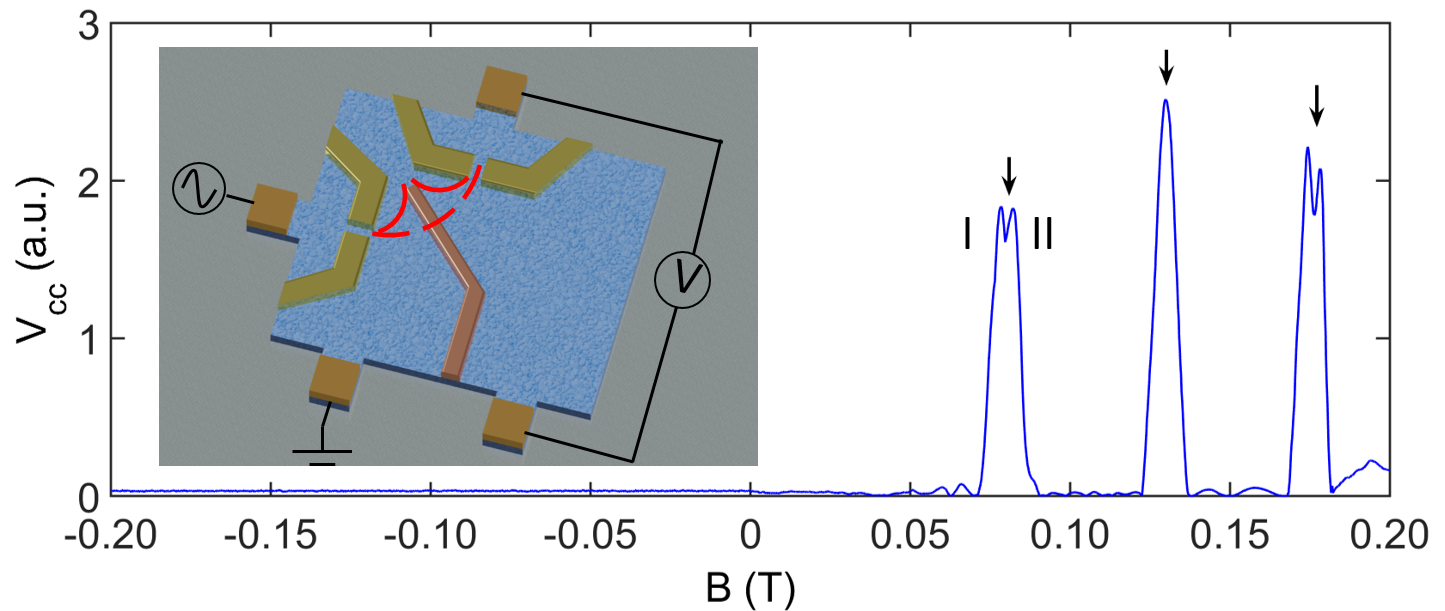} 
	\caption{The experiment setup and device characteristic. The periodic focusing peaks are well defined. Odd-numbered focusing peaks show splitting with emitter conductance set to G$_0$, the two subpeaks marked in the plot are referred to as peak I and peak II in the Letter. The inset shows a schematic outline of the experiment setup. The emitter, collector (shining yellow blocks) and beam splitter (brown block) are defined with metallic gates. Lithographically defined separation between the emitter and collector is 1.5 um (along diagonal direction), the width of beam splitter is 200 nm, and the QPC's length (width) is 400 nm (500 nm). The gap between the edge of emitter and collector is smaller (200 nm) than the width of the QPC, so when the emitter and collector are operational, this gap remains pinched off, thus, fully reflecting the incident electrons. The dashed and solid arc represent the orbit for the first and second focusing peaks, respectively. }           
	\label{fig:1}
\end{figure}

\begin{figure}

		\includegraphics[height=4.0in,width=2.8in]{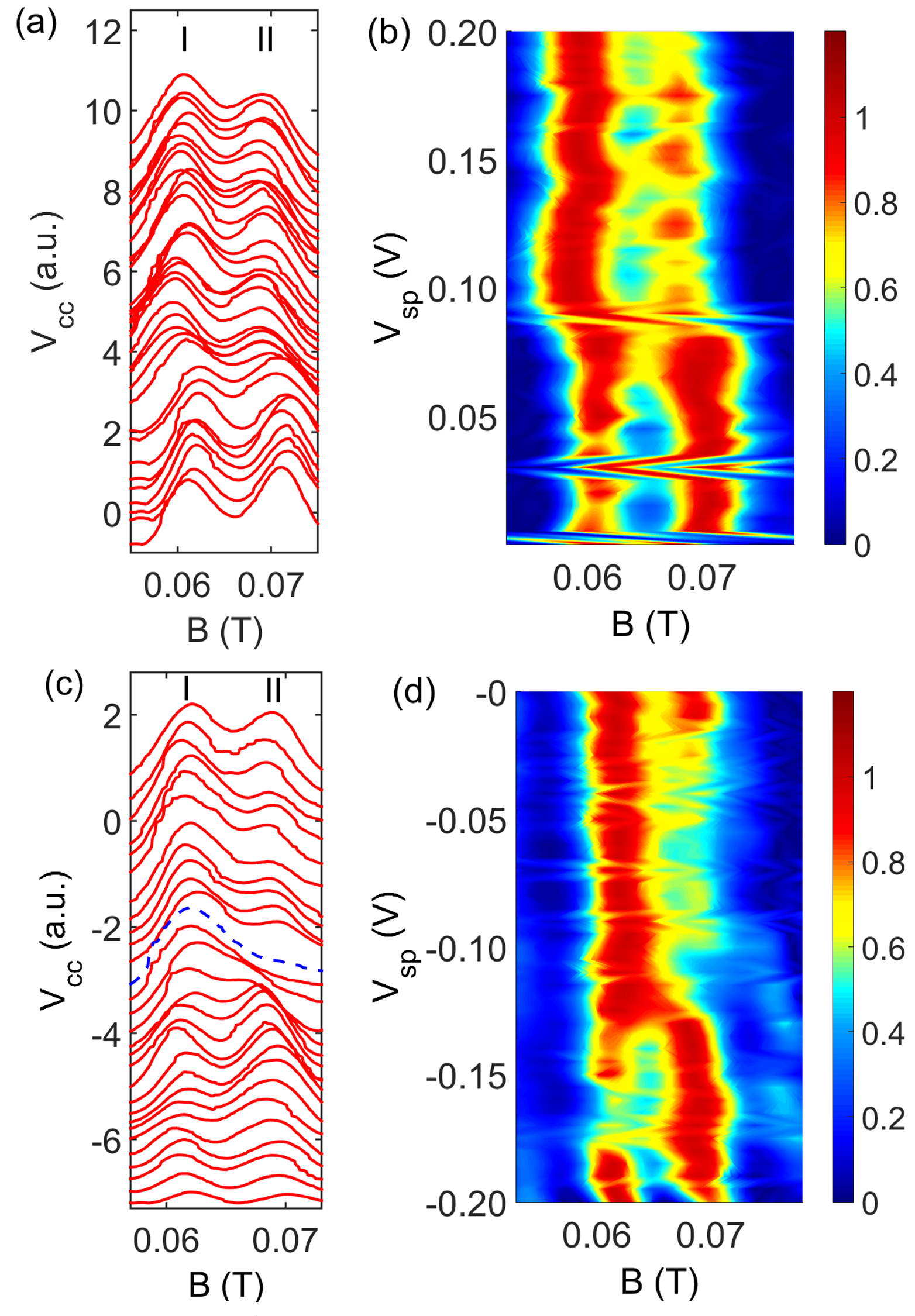}
	
	\caption{Focusing with beam splitter voltage. Both the emitter and collector were set to G$_0$. (a) The beam splitter voltage V$_{sp}$ was swept from 0.2 V (top trace) to 0 (bottom) trace. The data have been offset vertically for clarity. (b) Corresponding color map of data in (a). (c) The beam splitter voltage V$_{sp}$ was swept from 0 (top trace) to -0.2 V (bottom) trace. (d) Corresponding color map of data in (c). Each trace in (b) and (d) is normalized against its own maximum.   }           
	\label{fig:2}
\end{figure}

\begin{figure}
	
	\includegraphics[height=2.0in,width=3.6in]{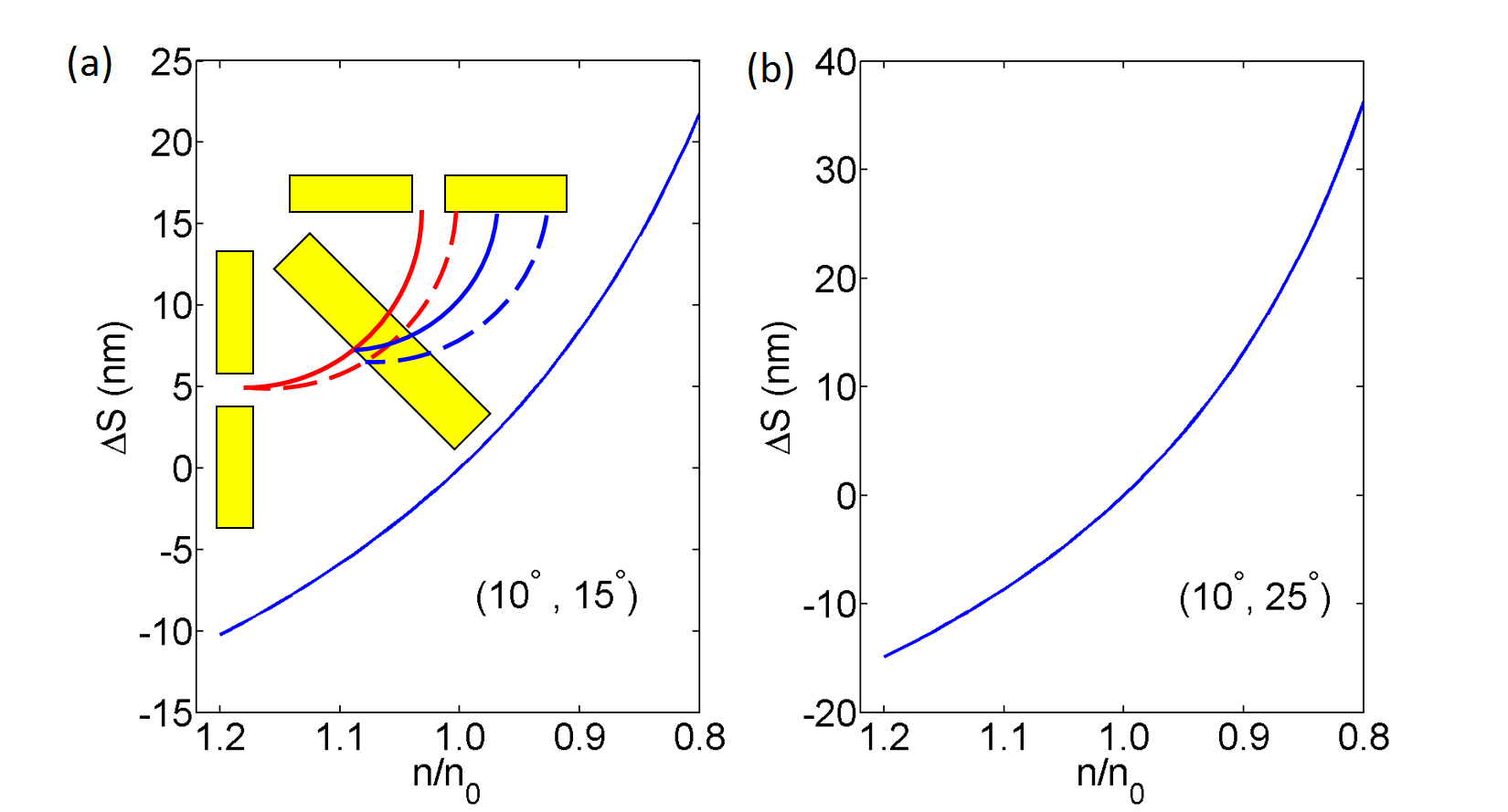}
	
	\caption{Electron refraction underneath the beam splitter. Calculated change in separation between the trajectories of two spin branches $\Delta$S as a function of $\frac{n}{n_0}$, for two sets of incident angles (10$^\circ$, 15$^\circ$) in (a) and (10$^\circ$, 25$^\circ$) in (b). Positive $\Delta S$ means the trajectories of the two spin branches get farther apart, while negative $\Delta S$ indicates the trajectories get closer. Inset in (a) shows a schematic for electron trajectories; when the beam splitter is grounded electrons take the red trajectories, the solid trace is spin-up electrons while the dashed trace is for spin-down electrons. When the beam-splitter is activated, electrons take the blue trajectories.  }    
	\label{fig:3}
\end{figure}

Experiment.---The devices studied in the present work were fabricated from a high mobility two-dimensional electron gas formed at the interface of GaAs/Al$_{0.33}$Ga$_{0.67}$As heterostructure. At 1.5 K, the measured electron density (mobility) was 1.80$\times$10$^{11} $cm$^{-2}$ (2.17$\times$10$^6 $cm$^2$V$^{-1}$s$^{-1}$); therefore, both the mean free path and phase coherence length (calculated from the Nyquist equation\cite{AAK82,PPP89}) were over 10 $\mu$m which was much larger than the electron propagation length. The experiments were performed in a cryo-free dilution refrigerator with a lattice temperature of 20 mK using a standard lock-in technique. 

Demonstrating spin amplification requires a direct measurement of spin polarization. Here we utilize a magnetofocusing scheme which is a well-established method in monitoring the spin polarization\cite{LCF17, YKT17,YKP17, LPR06,SC11,ARD06,YKP18}. We have used both the conventional linear geometry and an orthogonal geometry which was implemented in our previous work in studying the 1D spin gap\cite{YKT17,YKP17,YKP18}. In the present work, an additional inclined beam splitter was patterned 45$^\circ$ against both the emitter and collector as shown in inset of Fig.~\ref{fig:1} (characteristics of the QPCs and beam splitter are presented in Supplemental Material note 1).
     
In the presence of a small positive transverse magnetic field B$_{\perp}$ electrons are bent from the emitter to collector leading to focusing peaks periodic in B$_{\perp}$ with a periodicity\cite{HVH89} $B_{focus}=\frac{\sqrt{2}\hbar k_F}{eL}$. Here $\hbar$ is the reduced Planck constant, $k_F$ is the Fermi wave vector, $e$ is the elementary charge, $L$ is the separation between the emitter and collector along the diagonal direction and the prefactor $\sqrt{2}$ accounts for the orthogonal device geometry. The SOI introduces a difference in $k_{F}$ for the spin-up and spin-down branches; thus, the odd-numbered focusing peaks show a splitting\cite{LPR06,SC11,ARD06,YKT17,YKP17, LCF17,YKP18}. As shown in Fig.~\ref{fig:1}, the first and third focusing peaks split into two subpeaks, each representing a spin state, while the second focusing peak remains a single peak\cite{LPR06,SC11,ARD06,YKT17,YKP17, LCF17}. The amplitude of subpeaks $A_{\uparrow, \downarrow}$ is directly proportional to the population of a given spin branch $n_{\uparrow, \downarrow}$\cite{LPR06,SC11,ARD06}, which allows us to extract the spin polarization $p = |\frac{n_\uparrow - n_\downarrow}{n_\uparrow + n_\downarrow}| = |\frac{A_\uparrow - A_\downarrow}{A_\uparrow + A_\downarrow}|$. Figure~1 shows the representative focusing spectrum with both emitter and collector set to G$_0$ (G$_0$ = $\frac{2e^2}{h}$).

Result and discussion.---We have shown previously that in a magnetofocusing scheme the 1D emitter injects a stream of partially spin-polarized current into the 2D region whose spin polarization is visualized as the asymmetry between the magnitudes of the two subpeaks of the first focusing peak\cite{YKT17,YKP17, YKP18}. To achieve spin-polarization amplification we have used a capacitive beam splitter in the setup and measured the modulation of magnitudes of the subpeaks of the first focusing peak as a function of beam splitter voltage. We performed the focusing measurements by sweeping the beam splitter voltage V$_{sp}$ while maintaining the emitter conductance to G$_0$ (the results for 0.8G$_0$ and 1.2G$_0$ can be found in the Supplemental Material note 2).

Starting with positive $V_{sp}$, it is clear that both the subpeaks of first focusing peak were pronounced with $V_{sp}$ tuned from 0.2 V (top trace) to 0 (bottom trace) as shown in Fig.~\ref{fig:2}(a), however, the asymmetry in magnitudes of two subpeaks oscillated against $V_{sp}$, peak II was stronger than peak I at $V_{sp}$ =0.2 V whereas a reversal was observed at $V_{sp}$ = 0. We noted that the position of both subpeaks remained almost unaffected by the beam splitter [Fig.~\ref{fig:2}(b)].  

The focusing spectrum with negative $V_{sp}$ is considerably different from its positive counterpart as shown in Fig.~\ref{fig:2}(c). It is seen that the magnitude of both subpeaks attenuated rapidly with $V_{sp}$ swept from 0 (top trace) to - 0.2 V  (bottom trace) where both subpeaks showed a tendency to smear out (electrons underneath the beam splitter are fully depleted at $V_{sp}$ = -0.3 V). The weakening of both subpeaks is due to the reduction in transmission probability through the beam splitter on making V$_{sp}$ more negative. It is found that peak II gradually weakened first so that a highly asymmetric single peak was observed when $V_{sp}$ $\approx$ -0.12 V (highlighted by the dotted blue trace). On slightly reducing the beam splitter voltage to -0.13 V, both subpeaks reemerged. The fact that peak-II only disappeared for a narrow window of $V_{sp}$ is an indication of a relatively small energy difference between the two spin branches. Although the magnitude of two subpeaks was significantly affected by the beam splitter, the position of the subpeaks was insensitive to $V_{sp}$ as depicted in Fig.~\ref{fig:2}(d).

Origin of robustness of the peak position.---The splitting of odd-numbered focusing peaks is due to the SOI in the 2D region\cite{WPD03}, which has a general form 

\begin{equation} H_{so} = -\frac{\hbar}{4 m^{\ast 2} c^2} \vec{\sigma} \cdot \vec{p} \times \vec{E}\end{equation}

\noindent where $\vec{\sigma}$ is the Pauli matrix, $\vec{p}$ is the momentum, $\vec{E}$ is the electric field, $\hbar$ is the reduced Planck constant, $m^\ast$ is the electron effective mass, and \textit{c} is the speed of light. The dependence of $H_{so}$ on V$_{sp}$ is two-fold. First, $p \propto \sqrt{n_0 + CV_{sp}}$ ($n_0$ is the carrier density at zero gate voltage and \textit{C} is the effective capacitance). The metalization comprising the beam splitter is directly patterned on the cap layer of GaAs; therefore, the 2D electron density is sensitive to V$_{sp}$, which, in turn, results in an appreciable change in $p$. Second, $E \propto E_0 + a V_{sp}$ ($E_0$ is the built-in electric field due to quantum well asymmetry, while \textit{a} accounts for screening of the external electric field)\cite{SGE97}.  Therefore, it is expected that the focusing peak position should shift as a function of V$_{sp}$\cite{LCF17}.

In the present work, positive V$_{sp}$ should yield a stronger SOI (thus, a larger peak splitting) because it increases both the electron density and $E$ ($E_0$ is in parallel with external electric field induced by a positive gate voltage in the heterostructure utilized here). On the other hand, negative V$_{sp}$ should result in a smaller SOI (therefore, a smaller peak splitting). However, the experimental results indicate that the position of both the subpeaks almost remains unaffected by V$_{sp}$. Therefore, another mechanism might be playing a role here to compensate for the effect introduced by SOI.

One possible mechanism which may compensate for the peak shifting due to the varying SOI is the refraction of electrons as they enter the beam splitter\cite{GKD04}. The refraction follows Snell's law\cite{SHU90,SSB90,GKD04}. The calculated result suggests that the two spin branches get closer ($\Delta S$ $<$ 0) with positive V$_{sp}$ , while they are farther apart ($\Delta S$$>$0) with negative V$_{sp}$ as exemplified in Fig.~\ref{fig:3}. This trend is the opposite of the effect caused by SOI which results a change in spatial separation -10 nm $<$ $\Delta S$ $<$ 10 nm (positive sign in the separation change $\Delta S$ is for positive $V_{sp}$ corresponding to higher concentration and larger SOI); therefore, these two effects seem to cancel out each other.  It is necessary to mention that we cannot extract the change in SOI directly as the peak position almost remains unchanged\cite{LCF17}, so we adapt the value from a similar GaAs heterostructure\cite{SSE09}.

\begin{figure*}
           
           \includegraphics[height=4.5in,width=6.4in]{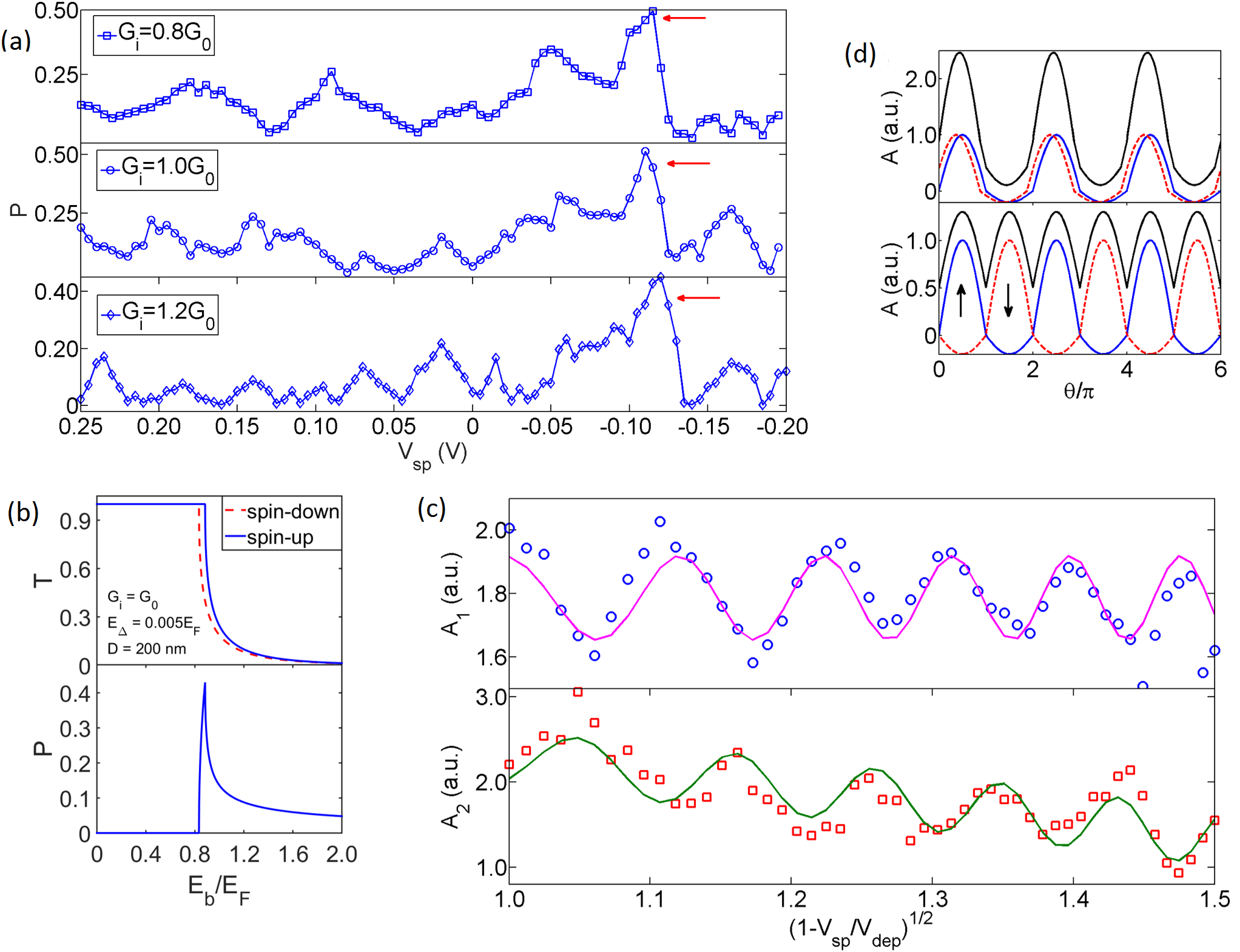}
      
\caption{Spin polarization with the beam splitter. (a) Spin polarization $P$ as a function of V$_{sp}$ for three difference emitter conductances. The overshoot in $P$ is highlighted by the arrows. (b) Upper plot shows the calculated transmission probability (T) through the splitter for spin-down (red) and spin-up (blue) electrons; lower plot is the corresponding polarization, E$_b$ is the barrier height and E$_F$ is Fermi energy. The width of beam splitter $D$ is set to 200 nm, the energy difference between the two spin branches $E_{\Delta}$ = 0.005E$_F$. (c) Amplitude of peak I (upper plot) and peak II (lower plot) as a function of $(1-\frac{V_{sp}}{V_{dep}})^{1/2}$ where $V_{dep}$ = -0.3 V is the pinched-off voltage of the beam splitter. The markers are raw data while the solid lines are sinusoidal fitting; a linear component has been added to the fitting for peak II to account for the background. (d) A graphical representation to illustrate the observed interference where the initial phase difference $\Delta \theta_0$ affects the period of $P$. The blue trace (red trace) is for spin-down (spin-up) branch, and their resultant is shown by black trace. A slower oscillation is obtained for small $\Delta \theta_0$ (upper panel) while a faster one for large $\Delta \theta_0$ (lower panel). }

	\label{fig:4}
	
\end{figure*}

Mechanisms for spin polarization amplification.---The spin polarization can be extracted from the asymmetry of two subpeaks \cite{AGC07,YKP17,YKT17} as  $P= \lvert \frac{A_1-A_2}{A_1+A_2} \rvert$, where $A_1$ and $A_2$ are the intensity of peak I and peak II, respectively and the result is shown in Fig.~\ref{fig:4}(a). It is clear that an overshoot in spin polarization is present in the regime  $- 0.10$ V $< V_{sp} < -0.13$ V  for all three emitter conductances, whereas a quasiperiodic oscillation is observed outside the regime (the oscillation is more pronounced in the ratio $\frac{A_1}{A_2}$ instead of $P$, as shown in Supplemental Material Fig.~7 ). With this approach,  a net spin polarization of around 50$\%$ can be achieved [see the central panel of Fig.~\ref{fig:4}(a)].

The overshoot can be understood via a model which simplifies the problem as tunneling through a 1D barrier. The rising potential barrier will block the spin-down electrons first (assuming they have lower energy) and leave the transmission probability $T$ of spin-up electrons almost unaffected until the barrier exceeds the energy of spin-up electrons (see Supplemental Material note 3). The calculated  $T$ for spin-down and spin-up electrons is shown in the upper panel of Fig.~\ref{fig:4}(b), and the difference in \textit{T} results in spin-polarization $P = |\frac{n_\uparrow - n_\downarrow}{n_\uparrow + n_\downarrow}| = |\frac{T_\uparrow - T_\downarrow}{T_\uparrow + T_\downarrow}| $ (with emitter set to G$_0$; lower panel) where the overshoot is reproduced. The highest spin polarization is achieved when $T$ for spin-up electrons starts reducing, corresponding to the blue dashed traces in Fig.~\ref{fig:2}(c). It is seen that despite the model being fairly simple, it accurately reproduces the highest achievable spin polarization of 42$\%$, while the experimentally observed polarization is  48$\%$. However, it is important to emphasise that 48$\%$ is not the upper limit of achievable spin polarization, as increasing the width of the beam splitter can significantly enhance the spin polarization; for instance, a spin polarization of $\sim$80$\%$ can be achieved when the width approaches 400 nm [see Supplemental Material Fig.6(a)]. The narrowness of the overshoot is a result of weak SOI for GaAs electron gas [see Supplemental Material Fig.6(b)].      

The potential barrier model predicts a nonoscillating spin polarization instead of the oscillatory one shown in Fig.~\ref{fig:4}(a). The oscillation at the negative V$_{sp}$ is enveloped by a change in the transmission probability and is relatively difficult to resolve; therefore, we analyze the positive V$_{sp}$ data to uncover the origin of the oscillation. There are two possible mechanisms which we consider, spin precession\cite{DD90,CHS15,LCF17} and quantum interference between electrons propagating underneath the beam splitter and those pass around without entering the gated region\cite{YSU91}. To trigger a spin precession with the observed large Larmor frequency, the estimated change rate of SOI with respect to V$_{sp}$ is of the order of 10$^{-11}$ eVm/V, which is 2 orders larger than reported elsewhere\cite{SSE09}; therefore, we suggest that spin precession is unlikely to be the reason behind the observed oscillation. In the quantum interference scenario, electrons passing through the beam splitter acquire  an extra phase owing to the change in $k_F$ (determined by $V_{sp}$) compared to electrons circumventing without entering the gated region. The oscillatory pattern should be periodic\cite{YSU91} against $(1-\frac{V_{sp}}{V_{dep}})^{1/2}$ where $V_{dep}$ = -0.3 V is the voltage when the beam splitter was fully pinched off, as shown in Fig.~\ref{fig:4}(c), with a periodicity of $\frac{2\pi}{k_F D}$, where $D$ is the width of the beam splitter. The periodicity yields a $D$ of 240 nm, which is consistent with the lithographically defined width of 200 nm. The difference in these values may be due to electrons following an arc trajectory instead of a straight line due to the cyclotron motion; thus, the effective width of the splitter experienced by the electrons is the length of the arc which is larger than 200 nm. 

It was noted that the oscillation in spin polarization with the emitter conductance set to 1.2G$_0$ was faster than that of 0.8G$_0$. Even though the periodicity of the individual spin branches is determined by the beam splitter voltage only, the initial phase difference $\Delta \theta_0$ between spin-up and spin-down branches can affect the periodicity of spin polarization as shown in Fig.~\ref{fig:4}(d). Theoretical work suggests\cite{YA02, SBV17} that the two spin branches may fluctuate in opposite directions when propagating through the 1D channel,  resulting in a phase change $\Delta \theta_0$  as they exit the 1D channel. It is important to emphasize that $\Delta \theta_0$ is determined by the emitter 1D channel only while the beam splitter does not affect $\Delta \theta_0$. The spin fluctuation (and so $\Delta \theta_0$) depends on the traversal time through the 1D channel\cite{SBV17} which is determined by the conductance accounting for why periodicity for 1.2G$_0$ differs from 0.8G$_0$.

Conclusion.---In conclusion, we demonstrate a non-spin-precession-based spin amplification in n-type GaAs utilizing a beam splitter patterned between the emitter and collector in a magnetofocusing configuration. The emitter injects two spin branches spatially separated into the 2D region, and the spin polarization can be accurately controlled and amplified by a capacitive  beam splitter. Such a system where the spin can be precisely controlled and amplified in addition to having a long spin-relaxation time is of interest for future technological schemes involving spin modulation. 

The work is funded by the Engineering and Physical Sciences Research Council, United Kingdom.

\end{document}